\documentclass{webofc}
\usepackage[varg]{txfonts}   
\usepackage{subcaption}

\newcommand{\beq}{\begin{equation}}
\newcommand{\eeq}{\end{equation}}
\begin{document}
\title{\vspace{-2.5cm} Study of a longitudinally expanding plasma
with the 2PI effective action}
%
%

\author{\firstname{François} \lastname{Gelis}\inst{1}\fnsep \and
        \firstname{Sigtryggur} \lastname{Hauksson}\inst{1}\fnsep\thanks{\email sigtryggur.hauksson@ipht.fr} }

\institute{Institut de Physique Th\'eorique,
  CEA/Saclay, Universit\'e Paris-Saclay,
91191 Gif sur Yvette, France 
          }

\abstract{%

A central question in heavy-ion collisions is how the initial far-from-equilibrium medium evolves and thermalizes while it undergoes a rapid longitudinal expansion. In this work we use the two-particle irreducible (2PI) effective action for the first time to consider this question, 
focusing on \(\phi^4\) scalar theory  truncated at three loops. We calculate the momentum distribution of quasiparticles in the medium and show that isotropization takes place. We furthermore consider the thermal mass of quasiparticles and the importance of number-changing processes. 
}
\maketitle
\section{Introduction}
\label{intro}


Heavy-ion collision experiments produce hot QCD matter. This matter has been shown to be well described by relativistic hydrodynamics. This raises the question of how a hydrodynamic description arises in a 
far-from-equilibrium QCD medium created in the collision of two ions. Current  state-of-the-art models  assume an initial classical regime, applicable due to high gluon occupancy at very early times, followed by a kinetic theory evolution of QCD quasiparticles \cite{review}.
Despite the appeal of this picture, it has some shortcomings. Firstly, an abrupt change from classical fields to quasiparticles makes it difficult to study e.g. the role of instabilities and non-thermal fixed points. Secondly, both frameworks make assumptions that limit their applicability across different energy scales. 
Thus a more unified and fundamental framework is needed.

\begin{figure}
\centering
\hfill
\begin{subfigure}{0.05\textwidth}
    \includegraphics[width=\textwidth]{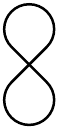}
\end{subfigure}
\hfill
\begin{subfigure}{0.15\textwidth}
    \includegraphics[width=\textwidth]{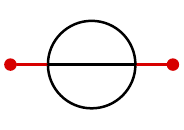}
\end{subfigure}
\hfill
\begin{subfigure}{0.08\textwidth}
    \includegraphics[width=\textwidth]{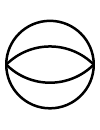}
\end{subfigure}
\hfill      
\caption{2PI bubble diagrams that contribute to the effective action \(\Gamma[\phi,D]\), truncated at three loops. The first diagram is called the tadpole diagram. The second diagram has two field insertions \(\phi\).}
\label{fig:Phi_diagrams}
\end{figure}

In this work we take the first step towards a unified  description of the initial stages of heavy-ion collisions. We use the two-particle irreducible (2PI) effective action \(\Gamma[ D]\) which 
depends on a resummed two-point function \(D(x,y)\).\footnote{In general \(\Gamma[\phi,D]\) where \(\phi\) is a one-point function. We set \(\phi =0\) in this work.} The effective action includes terms that describe free evolution and  2PI bubble diagrams that describe interactions \cite{2PI_review}. Without truncation the action is equivalent to the full quantum field theory. In practice one does a truncation leaving bubble diagrams up to three loops, see Fig. \ref{fig:Phi_diagrams}.  The 2PI effective action with this truncation contains both classical field theory and kinetic theory in their respective limits and thus encompasses the whole of the initial stages.
In this work we will apply the 2PI effective action to \(\phi^4\) scalar field theory with classical Lagrangian
\begin{align}
  {\cal L}\equiv \frac{1}{2}(\partial_\mu\phi)(\partial^\mu\phi)-\frac{m^2}{2}\phi^2 -\frac{g^2}{4!}\phi^4.
  \label{eq:lagrangian}
\end{align}
 This allows us to study isotropization in a simple context. 


\section{The 2PI effective action and longitudinal expansion}

The main novelty in this work is to use the 2PI action with a rapid longitudinal expansion as is found in heavy-ion collisions. (For an earlier proof-of-concept calculation see \cite{Hatta} and for an isotropic expansion in cosmology see \cite{Tranberg}.)
The relevant coordinate system is the Milne coordinates of proper time \(\tau\), rapidity \(\eta\) and position in the transverse plane \(\mathbf{x}_{\perp}\). We label the conjugate momenta to \(\eta,\mathbf{x}_{\perp}\) as \(\nu,\mathbf{p}_{\perp}\).
The equation of motion are expressed in terms of the statistical functions \(F(\tau, \tau^{\prime}; p_{\perp},\nu)\) and the spectral function \(\rho(\tau, \tau^{\prime}; p_{\perp},\nu)\).\footnote{There are two 2-point functions, \(F\)  and \(\rho\) because we work on the Keldysh-Schwinger contour. They are defined as \(F(x,y) = \frac{1}{2} \langle \{\phi(x),\phi(y) \}\rangle\) and \(\rho(x,y) = -i\langle [\phi(x),\phi(y) ]\rangle\).} The equation of motion for \(F\) is 
\begin{align}
&\Big[\partial_\tau^2+\tfrac{1}{\tau}\partial_\tau+M^2(\tau)+p_\perp^2+\tfrac{\nu^2}{\tau^2}\Big]\,F(\tau,\tau',p_\perp,\nu)\nonumber\\
&=
\int_{\tau_{\rm init}}^{\tau}d\tau''\tau''\;\Sigma_\rho(\tau,\tau'',p_\perp,\nu)\,F(\tau'',\tau',p_\perp,\nu)
+
\int_{\tau_{\rm init}}^{\tau'}d\tau''\tau''\;\Sigma_{_F}(\tau,\tau'',p_\perp,\nu)\,\rho(\tau'',\tau',p_\perp,\nu),
\label{eq:eom-F-p}
\end{align}
with a similar equation for \(\rho\). We have assumed homogeneity in the transverse plane and boost invariance. 
The left hand side of Eq. \eqref{eq:eom-F-p} contains the d'Alembertian \(\square = \partial_{\mu} \partial^{\mu}\) written in the coordinates \((\tau,p_{\perp},\nu)\) as well as an effective mass $M^2(\tau)$, defined as 
\begin{align}
\label{Eq:M2}
  M^2(\tau)\equiv m^2 +\tfrac{g^2}{2}\int \frac{d^2 p_\perp \,d\nu}{(2\pi)^3}\,\left[ F(\tau,\tau,p_\perp,\nu) - F_0(\tau,\tau,p_\perp,\nu) \right].
\end{align}
Here \(m^2\) is the vacuum mass. The right-hand side contains so-called memory integrals which describe scattering and which depend on the whole history of the system. They include self-energies \(\Sigma_{\rho}\) and \(\Sigma_F\) which come from the bubble diagrams in Fig. \ref{fig:Phi_diagrams} by cutting open one propagator. The detailed expression for the self-energies can be found in \cite{2PI_review}.  
In the current work we use a minimal form of renormalization where only power-law divergences in the tadpole are removed, see Eq. \eqref{Eq:M2}, where \(F_0\) is the bare propagator. For this reason we focus on observables that have little UV sensitivity. For further details on numerical implementation see \cite{RefSHFG}.

\section{Results}

\begin{figure}[htbp]
  \centering
  \includegraphics[width=0.6\textwidth]{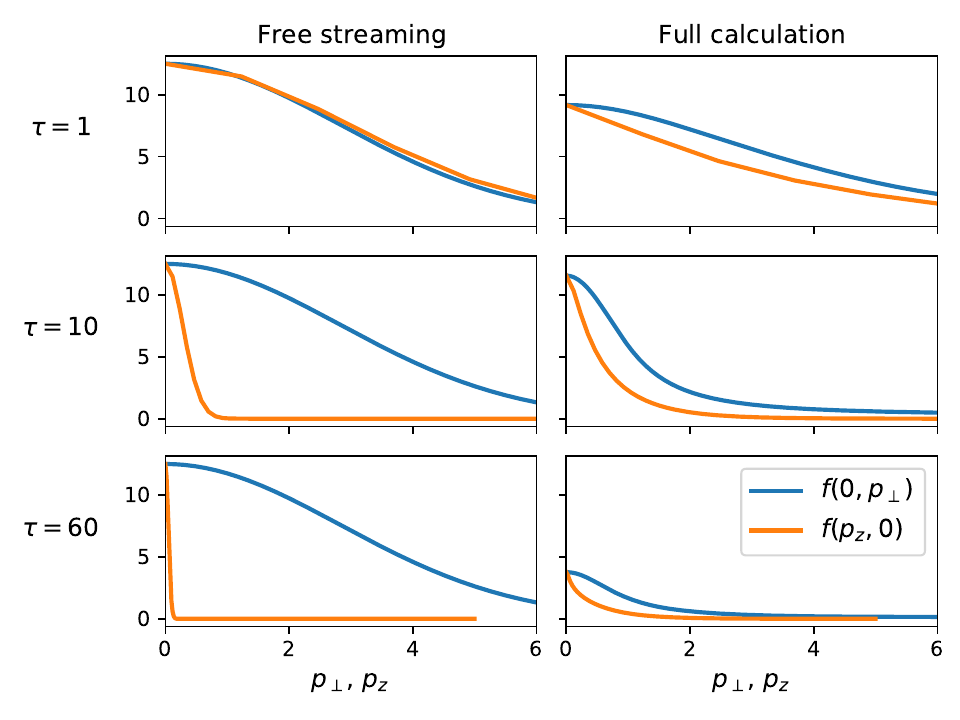}
  \caption{The occupation density \(f(p_z, p_{\perp})\) extracted from 2PI simulations. The left panel describes free streaming (\(g^4 = 0\)) while the right panel is a full calculation with \(g^4 = 500\). The full calculation shows isotropization as \(p_{\perp}\) and \(p_z\) have comparable magnitude at all times.}
  \label{fig:momdistr}
\end{figure}

\begin{figure}
    \centering
    \includegraphics[width=0.35\textwidth]{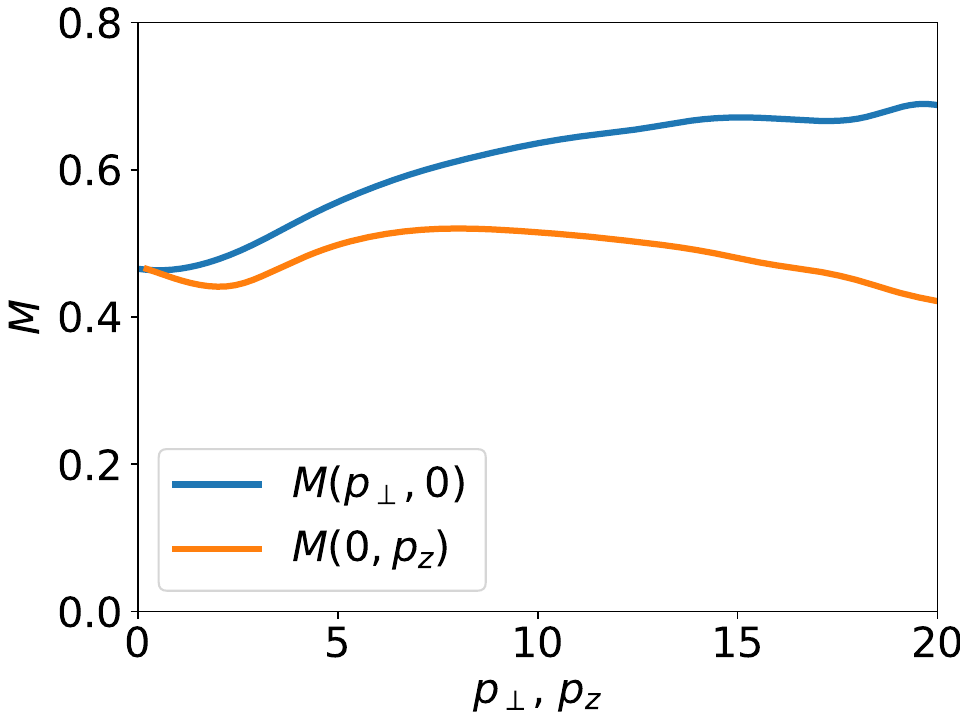}
    \caption{The effective mass \(M\) as a function of \(p_{\perp}\) and \(\nu\) at \(\tau = 10\). The effective mass has a moderate dependence on the momenta.}
    \label{fig:mass}
\end{figure}

\begin{figure}[htbp]
  \centering
  \includegraphics[width=0.5\textwidth]{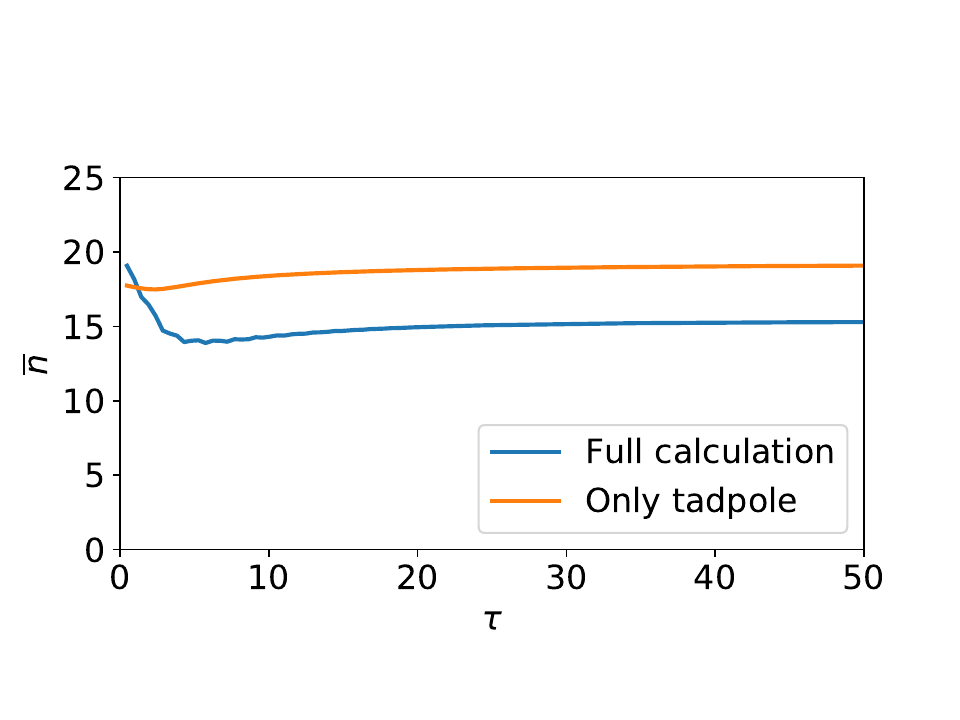}
  \caption{The evolution of number density with time for a full calculation and one which mean-field effects through the tadpole. The full calculation incorporates number-changing processes that account for a \(20\,\%\) reduction in number density.}
  \label{fig:number_density}
\end{figure}

In this work we initalize the system at time \(\tau=0.004\) with the free spectral function \(\rho\) and \(F\) given by
\beq
\label{Eq:initialF}
F(\tau_0, \tau_0; p_{\perp},\nu) = \left(\frac{1}{2} + f_0(p_{\perp},\nu)\right) \frac{\pi}{2} e^{-\pi \nu} \left| H_{i\nu}^{(1)}(m_T \tau_0)\right|^2
\eeq
where \(H^{(1)}_{i\nu}\) is a Hankel functions. Here \(m_T^2 = m^2 + p_{\perp}^2\) with vacuum mass \(m=2.5\).
This is the same expression as in vacuum except that we have an initial occupation density\footnote{This sets the typical momentum scale in our units as \(4.0\). For comparison the typical scale in the color-glass condensate is \(Q_s \approx 2.0\,\mathrm{GeV}\). This means that the time scale \(\tau \sim 1/Q_s = 0.1\,\mathrm{fm}\) corresponds to \(\tau \sim 1/4=0.25\) in our units.} 
\(
f_0(p_{\perp}, \nu) = e^{-p_{\perp}^2/\alpha^2} e^{-\nu^2/\beta^2}
\)
where \(\alpha = \beta = 4.0\).

To study isotropization microscopically we will extract an occupation density from the statistical function \(F\) at different times. We fit the full \(F\) from the 2PI simulation to a quasiparticle ansatz
\begin{align}
\label{Eq:F_quasiparticle}
  {F}_{\mathrm{quasiparticle}}(\tau,\tau';p_\perp,\nu)
  &=
  \frac{\pi (\tfrac{1}{2}+f(p_\perp,\nu;\tau))}{4}
  \left[
  H_{i\nu}^{(1)}(m_\perp \tau)H_{i\nu}^{(2)}(m_\perp \tau')+
  H_{i\nu}^{(2)}(m_\perp \tau)H_{i\nu}^{(1)}(m_\perp \tau')
  \right].
\end{align}
which has the same form as a  free propagator, except that there is a slowly varying occupation density \(f(p_\perp,\nu;\tau)\) and a slowly varying mass \(M(p_\perp. \nu;\tau)\) in \( m_\perp = \sqrt{m^2 + M^2(\tau) + p_{\perp}^2}\). 
In Fig. \ref{fig:momdistr} we show the extracted occupation density \(f(p_{\perp},p_{z}; \tau)\) at different times for \(g^4 = 500\) as well as for free streaming (\(g^4 = 0\)). (Here \(p_z = \nu/\tau\).) The full calculation shows clear signs of isotropization: the momenta \(p_{\perp}\) and \(p_z\) remain comparable at all times and the overall magnitude of the occupation density decreases to compensate for the larger extent of \(f\) in \(\nu\).
 This is unlike free streaming where the typical value of \(p_z\) falls like  \(1/\tau\) while \(p_{\perp}\) is constant.




The quasiparticle ansatz in Eq. \eqref{Eq:F_quasiparticle} contains an effective mass \(M(p_{\perp},\nu;\tau)\) that changes with time and which depends on the momentum. It includes non-perturbative corrections to the vaccum mass and thermal corrections. We show \(M\) in Fig. \ref{fig:mass} and see that it has moderate momentum dependence and is anisotropic. This information is not available in kinetic theory.

 Another important difference between a full 2PI calculation and kinetic theory is the possibility of number-changing processes in 2PI calculations. In kinetic theory all excitations are on shell so that \(1 \rightarrow 3\) and \(3 \rightarrow 1 \) processes are kinematically forbidden. This means that the  number density per unit rapidity and unit area in the transverse plane
\(
\overline{n}(\tau) = \int \frac{d^2 p_{\perp}d\nu}{(2\pi)^2} \; f(p_{\perp},\nu; \tau)
\)
is a conserved quantity. In the 2PI framework, excitatation can be off-shell and thus number-changing processes are allowed. In Fig. \ref{fig:number_density} we study the importance of this effect and see that in a full 2PI calculation \(\overline{n}\) changes substantially, especially at early times, giving an \(20\,\%\) overall reduction in the number of particles. This is expected because the initial conditions are overoccupied.


\section{Conclusion}

We have calculated the evolution of a longitudinally expanding medium for the first time using the two-particle irreducible (2PI) effective action. 
We see isotropization of the medium by looking at the evolution of the occupation density. We furthermore see that number-changing processes are important and that the thermal mass has a moderate momentum dependence.
The 2PI framework allows us to study many other aspects of non-equilibrium evolution in an expanding quantum field theory, such as isotropization in the stress-energy tensor \(T^{\mu\nu}\) and the decay of a background field \(\phi\) into quasiparticles.
We leave a more detailed discussion of these developments to future work.


\section*{Acknowledgements}
This work was granted access to the HPC resources of IDRIS under the allocation 2023-AD010514330 made by GENCI.

%
%
%

\end{document}